\documentclass[seceq,preprint]{ptptex}

\preprintnumber[3cm]{
YITP-08-79\\ September, 2008}

\title{
Double-Spinor Superstrings on Coset Superspaces%
}

\author{
Hiroshi  \textsc{Kunitomo}%
}

\inst{
Yukawa Institute for Theoretical Physics,
Kyoto University,\\
Kyoto 606-8502,
Japan
}

\abst{
The double-spinor formalism, proposed by Aisaka and Kazama,
provides a basis for the pure-spinor formalism, which allows 
manifestly super-Poincar\'e covariant quantization of superstrings.
We extend it to the case of backgrounds realized
by coset superspaces. 
A general method constructing reparametrization
invariant action is given using two concrete examples, 
the flat space-time and $AdS_5\times S^5$. 
We find that it is natural to double not only the spinor coordinates but
the whole superspace for describing double-spinor 
superstrings on such backgrounds.
The reparametrization invariant action has local symmetries compensating
those extra degrees of freedom, which guarantee
the equivalence to the Green-Schwarz formalism.
}

\begin{document}

\maketitle

\section{Introduction}

The pure spinor (PS) formalism is a promising method to quantize
superstrings in a manner that preserves the manifest super-Poincar\'e
covariance.\cite{B}  Defined as a free conformal field theory, 
it provides a calculable prescription to give general supersymmetric 
amplitudes in the flat ten-dimensional space-time.\cite{B2} 
The PS formalism can also be extended to the case of superstrings on 
the $AdS_p\times S^p$-type backgrounds\cite{BC,ADMO} and those obtained as
their Penrose limit.\cite{B3} It is essential to prove conformal invariance 
at the quantum level,\cite{B4} which cannot be realized with the Green-Schwarz (GS)
formalism. However, because of the lack of reparametrization invariant action,
the origins of various tools of the PS formalism, the pure spinor field, 
the BRST symmetry, the Feynman-like rules, etc., are still unclear.

To clarify this fundamental question, Aisaka and Kazama proposed 
the double spinor (DS) formalism based on a fundamental 
reparametrization invariant action from which one can derive the PS formalism.\cite{AK}
In the DS formalism, superstrings are propagating in an extended superspace,
$(X^a,\theta^\alpha_A,\tilde{\theta}^\alpha_A)$, obtained using doubling spinor coordinates.
The reparametrization invariant action has a novel local fermionic symmetry,
in addition to the conventional $\kappa$-symmetry, compensating
the extra spinor degrees of freedom $\tilde{\theta}^\alpha_A$.
It reduces to the GS action if the extra fermionic fields 
$\tilde{\theta}^\alpha_A$ are gauged away,
which yields the equivalence to the GS formalism.
The equivalence to the PS formalism is also proved in Ref.~\citen{AK},
which also gives an indirect proof of the equivalence between
the PS and GS formalisms.\footnote{
A similar proof is also given in Refs.~\citen{BM},~\citen{GG}.}
In this proof, the local fermionic symmetry plays 
an alternative role, an origin of the BRST symmetry. 
The pure spinor field $\lambda^\alpha_A$ is identified 
as a part of its ghost field.\cite{AK} 

The purpose of this paper is to extend the DS formalism to the backgrounds
realized by coset superspaces. 
Although the method is general and applicable to any coset superspace,
on which the GS superstring can be defined,
we illustrate it using two fundamental examples, 
flat ten-dimensional space-time and $AdS_5\times S^5$.
For general coset superspace, we find that it is natural to double 
the vector degrees of freedom as well as the spinor ones
as $(X^a,\tilde{X}^a,\theta_A^\alpha,\tilde{\theta}_A^\alpha)$.
The reparametrization invariant action of this \textit{extended} DS superstring 
also has local bosonic symmetry, in addition to the local fermionic symmetry,
compensating the extra degrees of freedom, $(\tilde{X}^a, \tilde{\theta}_A^\alpha)$. 
The action coincides with that of the GS formalism
if these extra fields $(\tilde{X}^a, \tilde{\theta}_A^\alpha)$ are gauged away.
This is an extension of the conventional DS formalism in the sense that 
the action reduces the conventional one after gauging away only the extra
bosonic field $\tilde{X}^a$ in the case of flat space-time.

The paper is organized as follows. In \S\ref{GS}, we briefly review the GS superstring
on flat space-time and $AdS_5\times S^5$ to fix our notations. The coset superspace
$SUSY(N=2)/SO(9,1)$ and $PSU(2,2|4)/(SO(4,1)\times SO(5))$ are introduced as
the target (super)space of the GS superstring on the flat space-time 
and $AdS_5\times S^5$, respectively. 
Given in \S\ref{DS} is the method to give the reparametrization
invariant action of the extended DS formalism.
The doubled superspace $(X^a,\tilde{X}^a,\theta^\alpha_A,\tilde{\theta}^\alpha_A)$ is naturally
introduced but the explicit form of the action can be obtained only in the $\tilde{X}^a=0$ gauge. 
The equivalence to the PS formalism is discussed in the final section. 
Appendix \ref{con} contains a summary of our conventions for spinors
and supergroups.

\section{Green-Schwarz superstring on the coset superspace}\label{GS}

In this section, we summarize the GS superstring on backgrounds 
realized as the coset superspace using two concrete examples, 
flat space-time and $AdS_5\times S^5$. 
We will extend them to the DS superstring in the next section.

\subsection{Flat background as a coset superspace}\label{eDS-flat}

The simplest example of the background realized as a coset superspace
is the flat background.
The GS superstring on the ten-dimensional flat space-time can be described as
a nonlinear sigma model on the coset superspace $SUSY(N=2)/SO(9,1)$.\cite{HM} 
Here, the $SUSY(N=2)$ $(SO(9,1))$ is the $N=2$ super-Poincar\'e (Lorentz) group
in ten dimensions, the details of which are summarized in Appendix~\ref{con}. 

This coset superspace can be parametrized by coordinates $(X^a,\theta^\alpha_A)$ as
\begin{subequations}\label{paraflat}
\begin{align}
&G(X,\theta)=g(X)\mathcal{G}(\theta),\label{group}\\
g(X)=&\exp(X^aP_a),\quad
\mathcal{G}(\theta)=\exp(\theta_A^\alpha Q^A_\alpha),
\end{align}
\end{subequations}
where $(X^a)^\dag=X^a,\ (\theta^\alpha_A)^\dag=\theta^\alpha_A$.
The structure of this coset superspace is exceptionally simple,
since $(P_a,Q^A_\alpha)$, frequently called broken generators, satisfy a closed subalgebra.
The nonlinear sigma model is described using the map from 
the world-sheet $\Sigma$, parametrized by  $\sigma^m$ with $ m=0,1$, 
to this coset superspace, $(X^a(\sigma^m),\theta^\alpha_A(\sigma^m))$.
The global supersymmetry
is nonlinearly realized on these fields by left multiplication as
\begin{equation}
 G(X,\theta)\rightarrow\mathcal{G}(\epsilon)G(X,\theta)=G(X',\theta'),
\end{equation}
or equivalently
\begin{equation}
 \delta X^a=-i\theta_A\bar{\gamma}^a\epsilon_A,\qquad
 \delta\theta_A^\alpha=\epsilon_A^\alpha,
\end{equation}
where $\epsilon_A^\alpha$ is a constant spinor parameter.

The left-invariant Maurer-Cartan (MC) one-form 
\begin{equation}
 G^{-1}dG=L^aP_a+L^\alpha_AQ^A_\alpha,
\end{equation}
is \textit{invariant} under the global supersymmetry,
where $L^a=L_m^ad\sigma^m$ and $L^\alpha_A=L^\alpha_{Am}d\sigma^m$ 
are the pullback of superspace one-forms
on the world-sheet using the sigma model map, which
can be easily calculated as
\begin{subequations}
\begin{align}
L^a_m=&\partial_m X^a-i\theta_A\bar{\gamma}^a\partial_m\theta_A,\\
L^\alpha_{Am}=&\partial_m\theta^\alpha_A.
\end{align}
\end{subequations}

The reparametrization invariant action is given as
\begin{subequations}\label{GSaction}
\begin{align}
I_{GS}=&I_K+I_{WZ},\\
I_K=&-\frac{1}{2}\int_{\Sigma} d^2\sigma \sqrt{-h}h^{mn}L^a_mL_{an},\label{kin}\\
I_{WZ}=&i\int_Ms^{AB}L^a\wedge L_A\wedge\bar{\gamma}_aL_B,
\end{align}
\end{subequations}
where $s^{11}=-s^{22}=1,\ s^{12}=s^{21}=0$, and $h_{mn}$ is the world-sheet metric. 
The Wess-Zumino action is integrated
over a three-dimensional manifold whose boundary is the world-sheet, $\partial M=\Sigma$.
This can be easily rewritten using the integral on the world-sheet as
\begin{equation}
I_{WZ}=
-i\int_\Sigma d^2\sigma \epsilon^{mn}(s^{AB}L^a_m\theta_A\bar{\gamma}_a\partial_n\theta_B
-i\theta_1\bar{\gamma}^a\partial_m\theta_1\theta_2\bar{\gamma}_a\partial_n\theta_2).\label{wz}
\end{equation}
The sum of (\ref{kin}) and (\ref{wz}) is nothing but the action of the GS superstring 
on the flat space-time.\cite{HM}

Due to the Wess-Zumino term, the action is invariant under the 
$\kappa$-symmetry defined as\cite{S}
\begin{subequations}\label{kappa}
\begin{align}
 \delta X^a=&i\theta_A\bar{\gamma}^a\delta\theta_A,\qquad
 \delta\theta_A=L^a_m\gamma_a\kappa_A^m,\\
 \delta(\sqrt{-h}h^{mn})=&-8i\sqrt{-h}P^{ml}_-\partial_l\theta_{1\alpha}\kappa^{\alpha n}_1
-8i\sqrt{-h}P^{ml}_+\partial_l\theta_{2\alpha}\kappa^{\alpha n}_2,
\end{align}
\end{subequations}
where
\begin{equation}
 P^{mn}_\pm=\frac{1}{2}\left(h^{mn}\pm\frac{\epsilon^{mn}}{\sqrt{-h}}\right),
\end{equation}
are the projection operators of the world-sheet vector. The parameters
$\kappa^m_{A\alpha}$ in (\ref{kappa}) are restricted to satisfy the constraints
\begin{equation}
P_-^{mn}\kappa_{1\alpha n}=\kappa^m_{1\alpha},\qquad 
P_+^{mn}\kappa_{2\alpha n}=\kappa^m_{2\alpha}.
\end{equation}
The $\kappa$-symmetry is crucial to reduce the fermionic degrees of freedom 
to the physical ones.

\subsection{$AdS_5\times S^5$ background}

Another important example of background, realized as a coset superspace, is
$AdS_5\times S^5$ with Ramond-Ramond (RR) flux obtained as 
a near-horizon limit of the D3-brane solution.\cite{Mal}
The GS superstring on this background is described as a nonlinear sigma model on
the coset superspace $PSU(2,2|4)/(SO(4,1)\times SO(5))$,\cite{MT,KR}
the details of which are given in Appendix~\ref{con}.
By using coordinates $(X^{\hat{a}},\theta_A^{\alpha\alpha'})$,
this is parametrized, in a similar way to (\ref{paraflat}),
as
\begin{subequations}\label{adsG}
\begin{align}
&G(X,\theta)=g(X)\mathcal{G}(\theta),\\
g(X)=&\exp(X^{\hat{a}}P_{\hat{a}}),\qquad
\mathcal{G}(\theta)=\exp(\theta^{\alpha\alpha'}_A Q^A_{\alpha\alpha'}),
\end{align}
\end{subequations}
where $(X^{\hat{a}})^\dag=X^{\hat{a}}$ and 
$ (\theta^{\beta\beta'}_A)^\dag{(\gamma^0)^\beta}_\alpha\delta^{\beta'}_{\alpha'}
=\theta_{\alpha\alpha'A}$.
The global supersymmetry is nonlinearly realized on these fields by left multiplication as
\begin{equation}
 G(X,\theta)\rightarrow\mathcal{G}(\epsilon)G(X,\theta)=G(X',\theta')H(\epsilon;X,\theta),
\end{equation}
where $\epsilon\ (=\epsilon_A^{\alpha\alpha'})$ is a global spinor parameter, and
$H(\epsilon;X,\theta)\in SO(4,1)\times SO(5)$ depends not only on
the parameter $\epsilon$, but also on the fields $(X^{\hat{a}},\theta_A^{\alpha\alpha'})$.

The left-invariant MC one-form is given as
\begin{equation}
 G^{-1}dG=L^{\hat{a}}P_{\hat{a}}+\frac{1}{2}L^{\hat{a}\hat{b}}M_{\hat{a}\hat{b}}
+L^{\alpha\alpha'}_AQ^A_{\alpha\alpha'},\label{MCads}
\end{equation}
where we use an abbreviation $L^{\hat{a}\hat{b}}M_{\hat{a}\hat{b}}\equiv L^{ab}M_{ab}+
L^{a'b'}M_{a'b'}$. Its components of broken generators, $(L^{\hat{a}},L^{\alpha\alpha'}_A)$, 
can be computed as\cite{KR}
\begin{subequations}\label{MCadscomp}
\begin{align}
 L^{\hat{a}}=&e^{\hat{a}}-4i\theta^A\hat{\gamma}^{\hat{a}}
\left(
\frac{\textrm{sh}^2(\mathcal{M}/2)}{\mathcal{M}^2}D\theta\right)^A,\\
L^{A\alpha}=&\left(
\frac{\textrm{sh}(\mathcal{M})}{\mathcal{M}}D\theta\right)^{A\alpha},
\end{align}
\end{subequations}
where 
\begin{align}
(\mathcal{M}^2)^{AB}=&
-\epsilon^{AC}\left((\gamma_a\theta_C)(\theta^B\gamma^a)
-(\gamma_{a'}\theta_C)(\theta^B\gamma^{a'})\right)
\nonumber\\
&\hspace{30mm}
+\frac{1}{2}\left((\gamma_{ab}\theta^A)(\theta_C\gamma^{ab})-
(\gamma_{a'b'}\theta^A)(\theta_C\gamma^{a'b'})\right)\epsilon^{CB},\\
D^{AB}=&\delta^{AB}\left(d+\frac{1}{4}\omega^{\hat{a}\hat{b}}\gamma_{\hat{a}\hat{b}}\right)
-\frac{i}{2}\epsilon^{AB}e^{\hat{a}}\hat{\gamma}_{\hat{a}}.
\end{align}
The one-form $e^{\hat{a}}$ $(\omega^{\hat{a}\hat{b}})$ is the pullback
of the viel-bein (spin-connection) of $AdS_5\times S^5$ space on the world-sheet.

The reparametrization invariant action on this background is given as
\begin{subequations}\label{adsgsaction}
\begin{align}
I_{GS}=&I_K+I_{WZ},\\
I_K=&-\frac{1}{2}\int_\Sigma d^2\sigma \sqrt{-h}h^{mn}L^{\hat{a}}_mL_{\hat{a}m},\\
I_{WZ}=&i\int_Ms^{AB}L^{\hat{a}}\wedge L_A\hat{\gamma}_{\hat{a}}\wedge L_B,
\end{align}
\end{subequations}
where the Wess-Zumino term can be written as the world-sheet integral,\cite{BBHZZ}
\begin{align}
 I_{WZ}=&-2\int_\Sigma L^1_{\alpha\alpha'}\wedge L^{\alpha\alpha'2},\nonumber\\
 =&-2\int_\Sigma d^2\sigma\epsilon^{mn}L^1_{\alpha\alpha'm}L^{\alpha\alpha'2}_n.
\end{align}
This action (\ref{adsgsaction}) is invariant under the $\kappa$-symmetry whose transformation laws
have simple forms on the linear combination
$(\delta x^{\hat{a}},\delta\vartheta_A^{\alpha\alpha'})$, obtained by replacing 
$(dX^{\hat{a}},d\theta_A^{\alpha\alpha'})$ in $(L^{\hat{a}},L_A^{\alpha\alpha'})$
with $(\delta X^{\hat{a}},\delta\theta_A^{\alpha\alpha'})$,
as\footnote{
At least perturbatively, one can solve them as the ones for the fundamental fields
$(X^{\hat{a}},\theta_A^{\alpha\alpha'})$, if necessary.} 
\begin{subequations}\label{kappa2}
\begin{align}
 \delta x^{\hat{a}}=0,\qquad&
 \delta\vartheta_A^{\alpha\alpha'}=L^a_m(\gamma_a\kappa_A^m)^{\alpha\alpha'}
-iL^{a'}_m(\gamma_{a'}\kappa_A^m)^{\alpha\alpha'},\\
 \delta(\sqrt{-h}h^{mn})=&-8i\sqrt{-h}\left(
P_-^{ml}L_{1\alpha\alpha' l}\kappa_1^{\alpha\alpha' n}
+P_+^{ml}L_{2\alpha\alpha' l}\kappa_2^{\alpha\alpha'n}
\right).
\end{align}
\end{subequations}
The parameters $\kappa_{Am}^{\alpha\alpha'}$ are restricted to be
\begin{equation}
P_-^{mn}\kappa^{\alpha\alpha'}_{1n}=\kappa^{\alpha\alpha'm}_1,\qquad 
P_+^{mn}\kappa^{\alpha\alpha'}_{2n}=\kappa^{\alpha\alpha'm}_2.
\end{equation}

\section{Double-spinor superstrings in coset superspace}\label{DS}

Now we extend the DS formalism to the backgrounds realized as coset superspaces.
Although the method is elucidated using two backgrounds explained in the previous
section, one can easily see that it is general and can be applicable to any consistent
background of GS formalism realized as coset superspaces.

\subsection{DS superstring in flat background}

Let us first reformulate, and slightly extend, the conventional DS superstring
in the flat space-time, in a way that is applicable to general backgrounds realized 
by coset superspaces. To begin with, we double the superspace coordinate fields 
$(X^a,\theta_A^\alpha)$ to 
$(X^a,\tilde{X}^a,\theta^\alpha_A,\tilde{\theta}^\alpha_A)$, and replace
the group-valued field $G(X,\theta)$ in (\ref{group}) by 
\begin{align}
\hat{G}(X,\tilde{X},\theta,\tilde{\theta})=&G(X,\theta)G^\dag(\tilde{X},\tilde{\theta}),
\nonumber\\
=&\exp\left(\left(\hat{X}^a-i(\theta_A\bar{\gamma}^a\tilde{\theta}_A)\right)P_a
+\Theta^\alpha_AQ^A_\alpha\right),\label{Ghat}
\end{align}
where $\hat{X}^a=X^a-\tilde{X}^a$ and 
$\Theta^\alpha_A=\theta^\alpha_A-\tilde{\theta}^\alpha_A$.\footnote{
Our convention is different by sign from the one in Ref.~\citen{AK}.}
It should be emphasized that the interaction terms between the conventional and extra spinor fields,
$-i(\theta_A\bar{\gamma}^a\tilde{\theta}_A)$,
which were introduced in a heuristic manner to realize a local fermionic symmetry in Ref.~\citen{AK},
naturally appeared in this doubled (group-valued) field $\hat{G}$ in (\ref{Ghat}). 
This is made possible by doubling not only the spinor coordinates $\tilde{\theta}_A$ 
but also the bosonic coordinates $\tilde{X}^a$, that is, the whole superspace, 
as a product of group-valued fields. 
It also makes it easy to extend the method to the case of general coset superspaces.

The global supersymmetry is, as in the case of the GS superstring,
realized by left-multiplication as
\begin{align}
 \hat{G}(X,\tilde{X},\theta,\tilde{\theta})
\rightarrow\mathcal{G}(\epsilon)\hat{G}(X,\tilde{X},\theta,\tilde{\theta})=&
\mathcal{G}(\epsilon)G(X,\theta)G(\tilde{X},\tilde{\theta})^\dag,\nonumber\\
=&G(X',\theta')G(\tilde{X},\tilde{\theta})^\dag=\hat{G}(X',\tilde{X},\theta',\tilde{\theta}),
\end{align}
which can be written as the transformations of the doubled superspace coordinate fields as
\begin{subequations}
\begin{alignat}{3}
 \delta X^a=&-i\theta_A\bar{\gamma}^a\epsilon_A,\qquad& \delta\theta_A^\alpha=&\epsilon_A^\alpha,\\
\delta\tilde{X}^a=&0,\qquad& \tilde{\theta}_A^\alpha=&0.\label{susy_flat}
\end{alignat}
\end{subequations}
The left-invariant MC one-form is defined using this doubled variable $\hat{G}$ as
\begin{equation}
\hat{G}^{-1}d\hat{G}=\hat{L}^aP_a+\hat{L}^\alpha_AQ^A_\alpha.
\end{equation}
We can easily obtain
\begin{subequations}
\begin{align}
\hat{L}^a=&d(\hat{X}^a-i\theta_A\bar{\gamma}^a\tilde{\theta}_A)
-i\Theta_A\bar{\gamma}^ad\Theta_A,\\
 \hat{L}^\alpha_A=&d\Theta_A^\alpha.
\end{align}
\end{subequations}

We define the reparametrization invariant action of the (extended) 
DS superstring by replacing $(L^a,L_A^\alpha)$,
in the action (\ref{GSaction}) of the GS superstring, 
with $(\hat{L}^a,\hat{L}_A^\alpha)$ as
\begin{subequations}\label{DSaction}
\begin{align}
I_{DS}=&I_K+I_{WZ},\\
I_K=&-\frac{1}{2}\int_{\Sigma} d^2\sigma \sqrt{-h}h^{mn}\hat{L}^a_m\hat{L}_{an},\\
I_{WZ}=&
-i\int_\Sigma d^2\sigma \epsilon^{mn}(s^{AB}\hat{L}^a_m\Theta_A\bar{\gamma}^a\partial_n\Theta_B
-i\Theta_1\bar{\gamma}^a\partial_m\Theta_1\Theta_2\bar{\gamma}_a\partial_n\Theta_2).
\end{align}
\end{subequations}
To study the local symmetries,
we start from the apparently trivial fact that
the doubled fundamental variable $\hat{G}$,
therefore the action (\ref{DSaction}),
is invariant under the local transformations,
\begin{subequations}\label{adslocalsym}
\begin{align}
G(X,\theta)'=&G(X,\theta)G(\xi,\chi),\\
G(\tilde{X},\tilde{\theta})'=&G(\tilde{X},\tilde{\theta})G(\xi,\chi),
\end{align}
\end{subequations}
where $\xi^a$ $(\chi_A^\alpha)$ is a local bosonic (fermionic) parameter.
This can be rewritten by the transformation of fields
$(X^a,\tilde{X}^a,\theta_A^\alpha,\tilde{\theta}_A^\alpha)$ as
\begin{subequations}
\begin{align}
 G(X,\theta)G(\xi,\chi)=&G(X',\theta'),\\
 G(\tilde{X},\tilde{\theta})G(\xi,\chi)
 =&G(\tilde{X}',\tilde{\theta}'),
\end{align}
\end{subequations}
which yields infinitesimal transformations
\begin{subequations}\label{localsym}
\begin{alignat}{3}
 \delta X^a=&\xi^a+i\theta_A\bar{\gamma}^a\chi_A,\qquad&
 \delta\theta_A^\alpha=&\chi_A^\alpha,\\
 \delta \tilde{X}^a=&\xi^a+i\tilde{\theta}_A\bar{\gamma}^a\chi_A,\qquad&
 \delta\tilde{\theta}_A^\alpha=&\chi_A^\alpha.
\end{alignat}
\end{subequations}
The action (\ref{DSaction}) reduces to the one of the GS superstring,
if the extra fields $(\tilde{X}^a,\tilde{\theta}^\alpha_A)$ are gauged away.
In addition to these local symmetries, the action (\ref{DSaction}) also has
the conventional $\kappa$-symmetry whose transformation law is
given  for the similar linear combination to (\ref{kappa2}),
obtained by replacing $(dX^a,d\tilde{X}^a,d\theta^\alpha_A,d\tilde{\theta}^\alpha_A)$
in $(\hat{L}^a,\hat{L}^\alpha_A)$
with $(\delta X^a,\delta\tilde{X}^a,\delta\theta^\alpha_A,\delta\tilde{\theta}^\alpha_A)$,
as
\begin{subequations}\label{kappa2ds}
\begin{align}
\delta x^a=&\delta(\hat{X}^a-i\theta_A\bar{\gamma}^a\tilde{\theta}_A)
-i\Theta_A\bar{\gamma}^a\delta\Theta_A=0,\label{kappa21}\\
\delta\vartheta_A^\alpha=&\delta\Theta_A^\alpha=
\hat{L}^a_m(\gamma_a\kappa_A^m)^\alpha,\label{kappa22}\\
 \delta(\sqrt{-h}h^{mn})=&-8i\sqrt{-h}P^{ml}_-\partial_l\Theta_{1\alpha}\kappa^{\alpha n}_1
-8i\sqrt{-h}P^{ml}_+\partial_l\Theta_{2\alpha}\kappa^{\alpha n}_2.
\end{align}
\end{subequations}
It should be noted that these transformations (\ref{kappa21}) and (\ref{kappa22})
cannot be solved uniquely using fundamental fields,
$(\delta X^a,\delta\tilde{X^a},\delta\theta_A^\alpha,\delta\tilde{\theta}_A^\alpha)$.
This is not a problem but a natural consequence of the local symmetries (\ref{localsym}).
More specifically, transformations of fundamental fields
$(\delta X^a,\delta\tilde{X^a},\delta\theta_A^\alpha,\delta\tilde{\theta}_A^\alpha)$,
producing the same combination (\ref{kappa21}) and (\ref{kappa22}), 
are related by the local symmetries (\ref{localsym}).
For example, let us consider two transformations,
\begin{subequations}
\begin{align}
\delta_1X^a=&-i\tilde{\theta}_A\bar{\gamma}^a\delta_1\theta_A
+i\Theta_A\bar{\gamma}^a\delta_1\theta_A,
\qquad \delta_1\tilde{X}^a=0,\\
\delta_1\theta_A^\alpha=&\hat{L}^a_m(\gamma_a\kappa_A^m)^\alpha,
\qquad \delta_1\tilde{\theta}_A^\alpha=0,
\end{align}
\end{subequations}
and
\begin{subequations}
\begin{align}
\delta_2X^a=&i\tilde{\theta}_A\bar{\gamma}^a\delta_2\tilde{\theta}_A,
\qquad \delta_2\tilde{X}^a=0,\\
\delta_2\theta_A^\alpha=&0,\qquad 
\delta_2\tilde{\theta}_A^\alpha=-\hat{L}^a_m(\gamma_a\kappa_A^m)^\alpha,
\end{align}
\end{subequations}
which give the same transformation laws as (\ref{kappa21}) and (\ref{kappa22}).
The difference between them,
\begin{subequations}
\begin{align}
\delta_1X^a-\delta_2X^a=&i\hat{L}^b_m\Theta_A\bar{\gamma}^a\gamma_b\kappa^m_A,\qquad
\delta_1\tilde{X}^a-\delta_2\tilde{X}^a=0,\\
\delta_1\theta_A^\alpha-\delta_2\theta_A^\alpha=&\hat{L}^a_m(\gamma_a\kappa^m_A)^\alpha,\qquad
\delta_1\tilde{\theta}_A^\alpha-\delta_2\tilde{\theta}_A^\alpha=\hat{L}^a_m(\gamma_a\kappa^m_A)^\alpha,
\end{align}
\end{subequations}
is a local transformation (\ref{localsym}) with parameters
\begin{equation}
 \xi^a=-i\theta_A\bar{\gamma}^a\chi_A,\qquad
 \chi_A^\alpha=\hat{L}^a_m(\gamma_a\kappa^m_A)^\alpha.
\end{equation}

Finally, the conventional DS formalism\cite{AK} is obtained by
setting $\tilde{X}^a=0$ using the local bosonic symmetry (\ref{localsym}).
The action (\ref{DSaction}) reduce to the original action,
which is invariant under the local fermionic symmetry\cite{AK}
\begin{subequations}
\begin{align}
 \delta X^a=&i\Theta_A\bar{\gamma}^a\chi_A,\\
 \delta\theta_A^\alpha=&\chi_A^\alpha,\qquad \delta\tilde{\theta}_A^\alpha=\chi_A^\alpha.
\end{align}
\end{subequations}
This residual symmetry, which keeps the gauge condition $\tilde{X}^a=0$
invariant, is obtained by combining the local fermionic transformation
in (\ref{localsym}) with the particular local bosonic transformation 
choosing the parameter $\xi^a=-i\theta_A\bar{\gamma}^a\chi_A$.

\subsection{DS superstring in $AdS_5\times S^5$} 

It is now straightforward to extend the DS formalism to the case of $AdS_5\times S^5$ background,
realized as the coset superspace $PSU(2,2|4)/(SO(4,1)\times SO(5))$.

First, we define $\hat{G}(X,\tilde{X},\theta,\tilde{\theta})=
G(X,\theta)G(\tilde{X},\tilde{\theta})^\dag$ using (\ref{adsG}), 
by doubling superspace coordinate fields $(X^{\hat{a}},\theta^{\alpha\alpha'}_A)$ to
$(X^{\hat{a}},\tilde{X}^{\hat{a}},\theta^{\alpha\alpha'}_A,\tilde{\theta}^{\alpha\alpha'}_A)$.
The left-invariant MC one-form has the same form as the one in the GS formalism (\ref{MCads}),
\begin{equation}\label{MCadsds}
\hat{G}^{-1}d\hat{G}=\hat{L}^{\hat{a}}P_{\hat{a}}
+\frac{1}{2}\hat{L}^{\hat{a}\hat{b}}M_{\hat{a}\hat{b}}
+\hat{L}^{\alpha\alpha'}_AQ^A_{\alpha\alpha'}.
\end{equation}
The global supersymmetry is still nonlinearly realized by left-multiplication as
\begin{align}\label{nlsusy}
 \hat{G}(X,\tilde{X},\theta,\tilde{\theta})
\rightarrow\mathcal{G}(\epsilon)\hat{G}(X,\tilde{X},\theta,\tilde{\theta})
=&G(X',\theta')H(\epsilon;X,\theta)G(\tilde{X},\tilde{\theta})^\dag,\nonumber\\
=&G(X',\theta')G(\tilde{X}',\tilde{\theta}')^\dag H(\epsilon;X,\theta),\nonumber\\
=&\hat{G}(X',\tilde{X}',\theta',\tilde{\theta}')H(\epsilon;X,\theta),
\end{align}
where $G(\tilde{X}',\tilde{\theta}')=
H(\epsilon;X,\theta)G(\tilde{X},\tilde{\theta})H(\epsilon;X,\theta)^\dag$. 
We should note that the extra fields $(\tilde{X}^{\hat{a}},\tilde{\theta}_A^{\alpha\alpha'})$ are
nontrivially transformed under the global supersymmetry contrary to the case of the flat space-time,
(\ref{susy_flat}). 
This implies that we cannot fix the conventional $\kappa$-symmetry keeping the supersymmetry manifest
by setting a half of $\tilde{\theta}_A$ to be zero as in Ref.~\citen{AK}. 
Thus, the relation to the PS formalism, which is manifestly super-Poincar\'e covariant, 
has to be much complicated.

One can define a reparametrization invariant action of the extended DS superstring,
by replacing the fundamental variable $G$ with this $\hat{G}$, in the action of the GS superstring 
(\ref{adsgsaction}), as
\begin{subequations}\label{ds2action}
\begin{align}
I_{DS}=&I_K+I_{WZ},\\
I_K=&-\frac{1}{2}\int_\Sigma d^2\sigma \sqrt{-h}h^{mn}\hat{L}^{\hat{a}}_m\hat{L}_{\hat{a}m},\\
I_{WZ}=&i\int_Ms^{AB}\hat{L}^{\hat{a}}\wedge \hat{L}_A\hat{\gamma}_{\hat{a}}\wedge \hat{L}_B,\nonumber\\
=&-2\int_\Sigma d^2\sigma\epsilon^{mn}\hat{L}^1_{\alpha\alpha'm}\hat{L}^{\alpha\alpha'2}_n.
\end{align}
\end{subequations}

The doubled fundamental variable $\hat{G}$ is invariant under the the local transformations,
\begin{subequations}\label{adslocalsym}
\begin{align}
G(X,\theta)'=&G(X,\theta)G(\xi,\chi),\\
G(\tilde{X},\tilde{\theta})'=&G(\tilde{X},\tilde{\theta})G(\xi,\chi),
\end{align}
\end{subequations}
where $(\xi^{\hat{a}},\chi_A^{\alpha\alpha'})$ are local parameters.
These transformations are again rewritten as the transformation on 
the superstring coordinate fields $(X^{\hat{a}},\tilde{X}^{\hat{a}},
\theta_A^{\alpha\alpha'},\tilde{\theta}_A^{\alpha\alpha'})$ in the following two steps.
At the beginning, the transformation (\ref{adslocalsym}) can be rewritten as
\begin{subequations}
\begin{align}
 G(X,\theta)G(\xi,\chi)=&G(X',\theta')H(X,\theta;\xi,\chi),\\
 G(\tilde{X},\tilde{\theta})G(\xi,\chi)
=&\tilde{H}(\tilde{X},\tilde{\theta};\xi,\chi)^\dag G(\tilde{X}',\tilde{\theta}'),
\end{align}
\end{subequations}
where $H(X,\theta;\xi,\chi)$ and $\tilde{H}(\tilde{X},\tilde{\theta};\xi,\chi)$
are $SO(4,1)\times SO(5)$ valued parameters depending on fields $(X^{\hat{a}},
\tilde{X}^{\hat{a}},\theta^{\alpha\alpha'}_A, \tilde{\theta}^{\alpha\alpha'}_A)$
as well as parameters $(\xi^{\hat{a}},\chi^{\alpha\alpha'}_A)$. 
This alternative form of transformation induces a nontrivial transformation on $\hat{G}$ as
\begin{align}
\hat{G}(X,\tilde{X},\theta,\tilde{\theta})'=&
G(X',\theta')H(X,\theta;\xi,\chi)
G(\tilde{X}',\tilde{\theta}')^\dag
\tilde{H}(\tilde{X},\tilde{\theta};\xi,\chi),\nonumber\\
=&\hat{G}(X',\tilde{X}'',\theta',\tilde{\theta}'')
\hat{H}(X,\tilde{X},\theta,\tilde{\theta};\xi,\chi),
\end{align}
where
\begin{subequations}
\begin{align}
\hat{G}(X',\tilde{X}'',\theta',\tilde{\theta}'')=&
G(X',\theta')G(\tilde{X}'',\tilde{\theta}'')^\dag,\\
G(\tilde{X}'',\tilde{\theta}'')=&H(X,\theta;\xi,\chi)
G(\tilde{X}',\tilde{\theta}')
H(X,\theta;\xi,\chi)^\dag,\\
\hat{H}(X,\tilde{X},\theta,\tilde{\theta};\xi,\chi)=&
H(X,\theta;\xi,\chi)\tilde{H}(\tilde{X},\tilde{\theta};\xi,\chi).
\end{align}
\end{subequations}
The action (\ref{ds2action}) is invariant under this transformation,
since the components of the broken generators, $(\hat{L}^{\hat{a}},\hat{L}_A)$,
are transformed covariantly.

The $\kappa$-transformation is defined on
the similar linear combination
$(\delta x^{\hat{a}},\delta\vartheta_A^{\alpha\alpha'})$ to (\ref{kappa2}), given by replacing 
$(dX^{\hat{a}},d\tilde{X}^{\hat{a}},
d\theta_A^{\alpha\alpha'},d\tilde{\theta}_A^{\alpha\alpha'})$ in $(L^{\hat{a}},L_A^{\alpha\alpha'})$
with $(\delta X^{\hat{a}},\delta\tilde{X}^{\hat{a}},
\delta\theta_A^{\alpha\alpha'},\delta\tilde{\theta}_A^{\alpha\alpha'})$, as
\begin{subequations}
\begin{align}
 \delta x^{\hat{a}}=0,\qquad&
 \delta\vartheta_A^{\alpha\alpha'}=\hat{L}^a_m(\gamma_a\kappa^m_A)^{\alpha\alpha'}
-i\hat{L}^{a'}_m(\gamma_{a'}\kappa^m_A)^{\alpha\alpha'},\\
 \delta(\sqrt{-h}h^{mn})=&-8i\sqrt{-h}\left(
P_-^{ml}\hat{L}_{1\alpha\alpha' l}\kappa^{\alpha\alpha'n}_1
+P_+^{ml}\hat{L}_{2\alpha\alpha' l}\kappa^{\alpha\alpha'n}_2
\right).
\end{align}
\end{subequations}
The transformation laws of the fundamental fields $(\delta X^{\hat{a}},\delta\tilde{X}^{\hat{a}},
\delta\theta_A^{\alpha\alpha'},\delta\tilde{\theta}_A^{\alpha\alpha'})$ cannot be uniquely determined,
but it is expected from the local symmetries.

In $\tilde{X}^a=0$ gauge, we can further find the explicit form of 
$(\hat{L}^{\hat{a}},\hat{L}_A^{\alpha\alpha'})$, 
in a similar way to the GS superstring\cite{MT,KR} as
\begin{subequations}
\begin{align}
\hat{L}^{\hat{a}}=&e^{\hat{a}}
-4i\theta^A\hat{\gamma}^{\hat{a}}\left(
\frac{\textrm{sh}^2(\mathcal{M}/2)}{\mathcal{M}^2}D\theta\right)^A
-4i\tilde{\theta}^A\hat{\gamma}^{\hat{a}}\left(
\frac{\textrm{sh}^2(\tilde{\mathcal{M}}/2)}{\tilde{\mathcal{M}}^2}D\tilde{\theta}\right)^A
\nonumber\\
&
+2i\tilde{\theta}^A\hat{\gamma}^{\hat{a}}\left(
\frac{\textrm{sh}(\tilde{\mathcal{M}})}{\tilde{\mathcal{M}}}
\frac{\textrm{sh}(\mathcal{M})}{\mathcal{M}}D\theta\right)^A
-8i\tilde{\theta}^A\hat{\gamma}^{\hat{a}}\left(
\frac{\textrm{sh}^2(\tilde{\mathcal{M}}/2)}{\tilde{\mathcal{M}}^2}
\hat{\mathcal{M}}^2\frac{\textrm{sh}^2(\mathcal{M}/2)}{\mathcal{M}^2}D\theta\right)^A,\\
\hat{L}^A=&\left(\textrm{ch}(\tilde{\mathcal{M}})\frac{\textrm{sh}(\mathcal{M})}
{\mathcal{M}}D\theta\right)^A
-\left(\frac{\textrm{sh}(\tilde{\mathcal{M}})}{\tilde{\mathcal{M}}}D\tilde{\theta}\right)^A
-2\left(\frac{\textrm{sh}(\tilde{\mathcal{M}})}{\tilde{\mathcal{M}}}
\hat{\mathcal{M}}^2\frac{\textrm{sh}^2(\mathcal{M}/2)}{\mathcal{M}^2}D\theta\right)^A,
\end{align}
\end{subequations}
where
\begin{subequations}
\begin{align}
(\tilde{\mathcal{M}})^{AB}=&
-\epsilon^{AC}\left(
(\gamma_a\tilde{\theta}_C)(\tilde{\theta}^B\gamma^a)
-(\gamma_{a'}\tilde{\theta}_C)(\tilde{\theta}^B\gamma^{a'})
\right)
\nonumber\\
&\hspace{30mm}
+\frac{1}{2}\left(
(\gamma_{ab}\tilde{\theta}^A)(\tilde{\theta}_C\gamma^{ab})
-(\gamma_{a'b'}\tilde{\theta}^A)(\tilde{\theta}_C\gamma^{a'b'})
\right)\epsilon^{CB},\\
(\hat{\mathcal{M}})^{AB}=&
-\epsilon^{AC}\left(
(\gamma_a\tilde{\theta}_C)(\theta^B\gamma^a)
-(\gamma_{a'}\tilde{\theta}_C)(\theta^B\gamma^{a'})
\right)
\nonumber\\
&\hspace{30mm}
+\frac{1}{2}\left(
(\gamma_{ab}\tilde{\theta}^A)(\theta_C\gamma^{ab})
-(\gamma_{a'b'}\tilde{\theta}^A)(\theta_C\gamma^{a'b'})
\right)\epsilon^{CB}.
\end{align}
\end{subequations}
One can easily see that they reduce to those of the GS superstring, (\ref{MCadscomp}), if we set
the extra fermionic field $\tilde{\theta}^{\alpha\alpha'}_A=0$.

\section{Discussion}\label{Dis}

In this paper, we extend the DS formalism for superstring in the flat space-time, 
proposed by Aisaka-Kazama,\cite{AK} to the case of backgrounds realized as coset superspaces. 
The action of the extended DS superstring is naturally defined on 
the doubled superspace $(X^a,\tilde{X}^a,\theta_A,\tilde{\theta}_A)$.
It has local symmetries compensating the extra degrees of freedom, 
$(\tilde{X}^a,\tilde{\theta}_A)$, and reduces to the GS action
if they are gauged away.
We explicitly construct the reparametrization invariant action 
for two fundamental examples, the flat space-time and $AdS_5\times S^5$.
It is easily seen that the method is general and applicable to any coset 
superspaces, which are consistent backgrounds of the GS superstring.

A remaining problem is to show the equivalence between
the extended DS formalism and the PS formalism.
It is difficult, however, since DS superstring cannot be exactly quantized
except for the case of the flat space-time.
We only point out here that the extended DS superstring reduces 
to the conventional one,\cite{AK} in the flat space-time limit.\footnote{
To be precise, it reduces to the extended DS superstring in the flat space-time,
given in \S\ref{eDS-flat}, which is trivially equivalent to the conventional one.}
On the other hand, there is an essential difference
between the DS superstrings in the flat and the $AdS_5\times S^5$ backgrounds.
In the latter case, the extra fields $(\tilde{X}^a,\tilde{\theta}_A)$ are nontrivially
transformed under the global supersymmetry, as given in (\ref{nlsusy}). 
Therefore, the manifest supersymmetry is broken if we fix the conventional 
$\kappa$-symmetry by setting a half of $\tilde{\theta}_A$ to be zero, 
as in Ref.~\citen{AK}.
This implies that the relation to the PS formalism, which has 
manifest super-Poincar\'e covariance, should be much complicated.

Finally, it is also interesting to find a geometric interpretation
of doubled superspace. The similar approaches\cite{II,Hull,BH},
doubling the (super) coordinates, may help the investigation.

\section*{Acknowledgements}
The work is supported in part by the Grant-in-Aid for Scientific Research 
No.\#19540284 and the Grant-in-Aid for the Global COE Program \lq\lq The Next Generation
of Physics, Spun from Universality and Emergence,'' from the Ministry of Education,
Culture, Sports, Science and Technology (MEXT) of Japan.

\newpage

\appendix
\section{Conventions for Spinors and Supergroups}\label{con} 

\subsection{$N=2$ super-Poincar\'e algebra in ten dimensions}

For $SO(9,1)$ gamma matrices,
\begin{subequations}
\begin{align}
\{\Gamma^a,\Gamma^b\}=&2\eta^{ab},\qquad (a,b=0,1,\cdots,9)\\
 (\Gamma^a)^T=&-C\Gamma^aC^{-1},\qquad C^T=-C,
\end{align}
\end{subequations}
where $\eta^{ab}=\textrm{diag}(-1,1,\cdots,1)$, we use a chiral Majorana 
representation
\begin{equation}
\Psi=
\begin{pmatrix}
 \psi^\alpha\\
 \chi_\alpha
\end{pmatrix},\qquad
 C=
\begin{pmatrix}
 0 & 1 \\
-1 & 0
\end{pmatrix},\qquad
\Gamma^a=
\begin{pmatrix}
 0 & (\gamma^a)^{\alpha\beta}\\
(\bar{\gamma}^a)_{\alpha\beta} & 0
\end{pmatrix},
\end{equation}
where
$(\gamma^a)^{\alpha\beta}=(\gamma^a)^{\beta\alpha}$ and 
$(\bar{\gamma}^a)_{\alpha\beta}=(\bar{\gamma}^a)_{\beta\alpha}$.

The $N=2$ super-Poincar\'e group in ten dimensions is generated 
by anti-hermitian formal generators $(P_a,M_{ab},Q_\alpha^A)$
satisfying
\begin{subequations}
\begin{align}
[M_{ab},M_{cd}]=&
\eta_{bc}M_{ad}-\eta_{bd}M_{ac}+\eta_{ad}M_{bc}-\eta_{ac}M_{bd},\\
[M_{ab},P_c]=&\eta_{bc}P_a-\eta_{ac}P_b,\qquad
[M_{ab},Q^A_{\alpha}]=\frac{1}{2}Q^A_\beta(\gamma_{ab})^{\beta\alpha} ,\\
&\{Q^A_\alpha, Q^B_\beta\}=-2i\delta^{AB}(\bar{\gamma}^a)_{\alpha\beta}P_a.
\end{align}
\end{subequations}
The subgroup $SO(9,1)$ is the ten-dimensional Lorentz group generated by $M_{ab}$.

\subsection{$PSU(2,2|4)$}

For $SO(4,1)$ gamma matrices, we use a convention
\begin{subequations}
\begin{align}
 \{\gamma^a,\gamma^b\}=&2\eta^{ab},\qquad (a,b=0,1,\cdots,4)\\
 (\gamma^a)^T=&C\gamma^aC^{-1},\qquad C^T=-C.
\end{align}
\end{subequations}
The identity $i\gamma^0\gamma^1\cdots\gamma^4=1$ holds.
Similarly, $SO(5)$ gamma matrices satisfy
\begin{subequations}
\begin{align}
 \{\gamma^{a'},\gamma^{b'}\}=&2\delta^{a'b'},\qquad (a',b'=5,6,\cdots,9)\\
 (\gamma^{a'})^T=&C'\gamma^{a'}C'^{-1},\qquad C^T=-C,
\end{align}
\end{subequations}
and the identity $\gamma^5\cdots\gamma^9=1$ holds.
It is useful to define $\hat{\gamma}^a=\gamma^a$ and $\hat{\gamma}^{a'}=i\gamma^{a'}$.

The supergroup $PSU(2,2|4)$ is generated by formal generators
$(P_{\hat{a}}, M_{\hat{a}\hat{b}},Q_{\alpha\alpha'}^A)$ satisfying hermiticity conditions
$(P_{\hat{a}})^\dag=-P_{\hat{a}},\ (M_{\hat{a}\hat{b}})^\dag=-M_{\hat{a}\hat{b}}$ and 
$(Q^{\beta\beta'A})^\dag{(\gamma^0)^\beta}_\alpha\delta^{\beta'}_{\alpha'}=-Q_{\beta\beta'}^A$, 
where we define
\begin{subequations}
\begin{align}
Q_{\alpha\alpha'}^A=&Q^{\beta\beta'A}C_{\beta\alpha}C'_{\beta'\alpha'},\\
Q^{\alpha\alpha'A}=&Q_{\beta\beta'}^A(C^{-1})^{\beta\alpha}(C'^{-1})^{\beta'\alpha'}.
\end{align}
\end{subequations}
We also use the abbreviated notations
$P_{\hat{a}}=(P_a,P_{a'})$ and $M_{\hat{a}\hat{b}}=(M_{ab},M_{a'b'})$.

These generators satisfy $psu(2,2|4)$ superalgebra given by
\begin{subequations}
\begin{align}
 [P_a,P_b]=&M_{ab},
\hspace{35mm} 
[P_{a'},P_{b'}]=-M_{a'b'},\\
[P_a,M_{bc}]=&\eta_{ab}P_c-\eta_{ac}P_b,
\hspace{15mm}
[P_{a'},M_{b'c'}]=\delta_{a'b'}P_{c'}-\delta_{a'c'}P_{b'},\\
[Q_{\alpha\alpha'}^A,P_a]=&
-\frac{i}{2}\epsilon_{AB}Q^B_{\beta\alpha'}{(\gamma_a)^\beta}_\alpha,
\hspace{5mm}
[Q_{\alpha\alpha'}^A,P_{a'}]=
\frac{1}{2}\epsilon_{AB}Q^B_{\alpha\beta'}{(\gamma_{a'})^{\beta'}}_{\alpha'},\\
[Q_{\alpha\alpha'}^A,M_{ab}]=&-\frac{1}{2}Q^A_{\beta\alpha'}{(\gamma_{ab})^\beta}_\alpha,
\hspace{7mm}
[Q_{\alpha\alpha'}^A,M_{a'b'}]=-\frac{1}{2}Q^A_{\alpha\beta'}{(\gamma_{a'b'})^{\beta'}}_{\alpha'},\\
[M_{ab},M_{cd}]=&\eta_{bc}M_{ad}-\eta_{bd}M_{ac}+\eta_{ad}M_{bc}-\eta_{ac}M_{bd},\\
 [M_{a'b'},M_{c'd'}]=&\delta_{b'c'}M_{a'd'}-\delta_{b'd'}M_{a'c'}+\delta_{a'd'}M_{b'c'}
-\delta_{a'c'}M_{b'd'},\\
\{Q^A_{\alpha\alpha'},Q^B_{\beta\beta'}\}=&\delta^{AB}
\left(
-2i(C\gamma^a)_{\alpha\beta}C'_{\alpha'\beta'}P_a
+2C_{\alpha\beta}(C'\gamma^{a'})_{\alpha'\beta'}P_{a'}
\right)
\nonumber\\
&\hspace{25mm}
+\epsilon^{AB}\left(
(C\gamma^{ab})_{\alpha\beta}C'_{\alpha'\beta'}M_{ab}
-C_{\alpha\beta}(C'\gamma^{a'b'})_{\alpha'\beta'}M_{a'b'}\right).
\end{align}
\end{subequations}
The subgroup $SO(4,1)\times SO(5)$
is generated by $M_{\hat{a}\hat{b}}$.

\end{document}